# Optimum Launch Power in Multiband Systems

Yanchao Jiang[(1)], Fabrizio Forghieri[(2)], Stefano Piciaccia[(2)], Gabriella Bosco[(1)], Pierluigi Poggiolini[(1)]

[(1)] DET, Politecnico di Torino, C.so Duca Abruzzi 24, 10129, Torino, Italy, yanchao.jiang@polito.it
[(2)] CISCO Photonics Italy srl, via Santa Maria Molgora 48/C, 20871, Vimercate (MB), Italy

**Abstract** *We investigate the residual throughput penalty due to ISRS, after power-optimization, in multi-band systems. We show it to be mild. We also revisit the launch power optimization "3-dB rule". We find that using it is possible but not advisable due to increased GSNR non-uniformity.* ©2024 The Author(s)

## Introduction

Many technologies are currently competing in the quest for increasing the throughput of optical links. Multiband is one of them. The extension from C to C+L is already commercially available and is enjoying substantial success. Research is focusing on adding further bands, primarily S but also E, U and even O. In this paper we consider medium-to-long-haul systems, about 300km to 1000km. We focus on C+L+S as primary option, with C+L+S+E also considered for the shorter end of the length range.

One prominent feature of multiband systems is the presence of strong Inter-Channel Raman Scattering (ISRS). It drastically increases the loss experienced by higher-frequency channels and decreases that of lower-frequency channels. It may amount to several dB and must be taken into account in the design of multiband systems. Specifically, launch power per channel must be carefully optimized. Up to now, an often-used criterion for the optimization of launch power has been to aim for each channel accumulating a noise power due to non-linear interference (NLI) equal to half that of ASE, that is:

$$P_{\mathrm{NLI}} = \tfrac{1}{2} P_{\mathrm{ASE}} \qquad (1)$$

This is sometimes called the "3-dB rule". The optimality of this criterion was proved first in [1], [2]. It was then shown to be enforceable span-by-span in [3]. Recently, Eq. (1) has been discussed in the context of multiband systems in [4]. In this paper we first review (1) and recall the assumptions the rule was based on.

Then we discuss launch power optimization in multiband systems in the presence of ISRS. We show that: the resulting optimum launch power leads to channel propagation regimes significantly different from (1); such optimization can reduce the throughput penalty due to ISRS to a few percentage points. We also show that, if the 3-dB rule is nonetheless enforced, the additional penalty is surprisingly mild, approximately 3% of throughput. However, other drawbacks show up, which we discuss in the paper.

The reported optimization results were obtained using a fast EGN closed-form-model (CFM), which was presented in [5] and extensively experimentally validated in [6]. They are in broad agreement with the findings of [4]. However, at optimum launch power we find rather different resulting propagation regimes for low and high frequency channels. We address this discrepancy when discussing the results.

## The "3-dB rule"

Optical system performance is governed by the GSNR at the Rx, defined as:

$$\mathrm{GSNR} = P_{\mathrm{ch}} / (P_{\mathrm{ASE}} + P_{\mathrm{NLI}}) \qquad (2)$$

Circa 2010, it was found that in uncompensated coherent links NLI scaled according to:

$$P_{\mathrm{NLI}} = \eta \cdot P_{\mathrm{ch}}^3 \qquad (3)$$

where $\eta$ depends on all link parameters, but *not* on launch power. By substituting (3) into (2) and then looking for the value of $P_{\mathrm{ch}}$ which maximizes (2), the optimum launch power is found to be:

$$P_{\mathrm{ch}_{opt}} = \sqrt[3]{P_{\mathrm{ASE}}/2\eta} \qquad (4)$$

Eq. (4) allows to obtain three key results. First, using (4) in (3) the "3-dB rule" (1) is directly found. Then, using (4) in (2) the corresponding max GSNR is derived:

$$\mathrm{GSNR}_{\max} = P_{\mathrm{ch}_{opt}} / \left(\tfrac{3}{2} P_{\mathrm{ASE}}\right) \qquad (5)$$

Finally, using the definitions:

$$\mathrm{OSNR} = P_{\mathrm{ch}} / P_{\mathrm{ASE}} \qquad (6)$$
$$\mathrm{GSNR}_{\mathrm{NLI}} = P_{\mathrm{ch}} / P_{\mathrm{NLI}} \qquad (7)$$

with (4), it turns out that the ratio between $\mathrm{GSNR}_{\mathrm{NLI}}$ and OSNR, at the optimum launch power, is also 3dB. In fact, this ratio can be used as a definition of the rule alternative to (1).

Eqs. (1)-(7) were originally derived under the assumption of all identical, equally spaced channels, and identical spans with frequency-independent parameters. However, over the years, this rule has proved to work reasonably well in much more general conditions [7-9]. This has led to the system optimization practice of trying to approach the 3-dB rule for each channel at each span. However, as we will show in the following, the 3-dB rule has limitations and breaks down when ISRS becomes significant.

## 1000 km C+L+S systems

We first look at a 1000km C+L+S link comprising

10 spans of 100km each of SMF. The link is derived from the digital twin of an actual 5-span experimental testbed, with average span loss of 22.5 dB. Each individual fiber of the system was characterized as for loss, dispersion, non-linearity coefficient and Raman gain profile, across the L, C and S bands. The details can be found in [6]. To achieve 1000 km, we repeat the 5-span link twice. The band boundaries are as follows: L-band 184.50 to 190.35; C-band 190.75 to 196.60; S-band 197.00 to 202.85 (5.85 THz per band). Doped-Fiber-Amplifiers (DFAs) were assumed with 6dB noise-figure in L- and S-band and 5dB in C-band. The WDM signal consisted of 50 channels in each band, with symbol rate 100 GBaud, roll-off 0.1 and spacing 118.75 GHz. Modulation was assumed Gaussian-shaped.

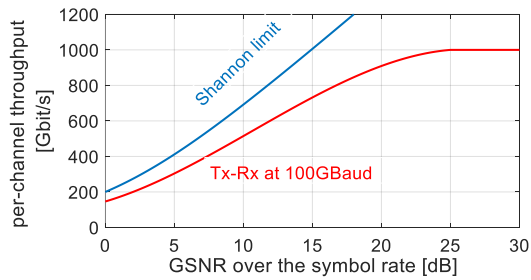

**Fig. 1:** Per-channel throughput vs. GSNR. Blue: Shannon limit; Red: high-performance Tx-Rx pair at 100 GBaud.

Throughput was found for each channel by calculating the GSNR (using the EGN CFM) and then resorting to the red curve shown in Fig.1, which is representative of top-performance commercial transponders operated at 100 GBaud. The launch power spectrum was described using a cubic polynomial in each band, for a total of 12 free parameters. The objective function used to optimize these 12 parameters was the *overall system throughput*.

The results with **ISRS off** are shown in Fig.2(a). Note that, across all figures, markers were computed with the numerically-integrated EGN model, to confirm the accuracy of the CFM. The optimum launch power per channel is mostly flat, about 5 dBm per channel, dipping slightly in the C-band. The corresponding maximized total net throughput was 96.50 Tb/s. Interestingly, the gap between GSNR$_{NLI}$ and OSNR is very close to 3-dB throughout the whole WDM spectrum, despite GSNR ranging from 14dB in the L-band down to 10.7dB in the S-band. This "emergence" of approximately the 3-dB rule from the system throughput maximization is a remarkable result, since many aspects of this system are frequency-dependent, such as loss, dispersion, non-linearity coefficient and noise figure. The slight deviations from exactly 3-dB are likely due to the different DFA noise figures in C band vs. S and L bands.

We then turned **ISRS on**. A very different picture emerges in Fig.2(b). The launch power maximizing throughput is no longer flat and now swings from -3.5 dBm at about 187 THz to 9.6 dBm at the top of the S band (about 203 THz). The GSNR too has much greater variability, ranging from 15 dB in L band to only 7 dB at the top of the S-band. Most notably, the 3-dB rule is broken. In the L-band, *propagation is essentially linear*, with GSNR$_{NLI}$/OSNR being far away from 3 dB, reaching up to 10 dB. In the S-band, such ratio is instead *lower* than 3dB, about 2.0 dB on average, and less at high frequency. The C-band shows a transition between the two regimes.

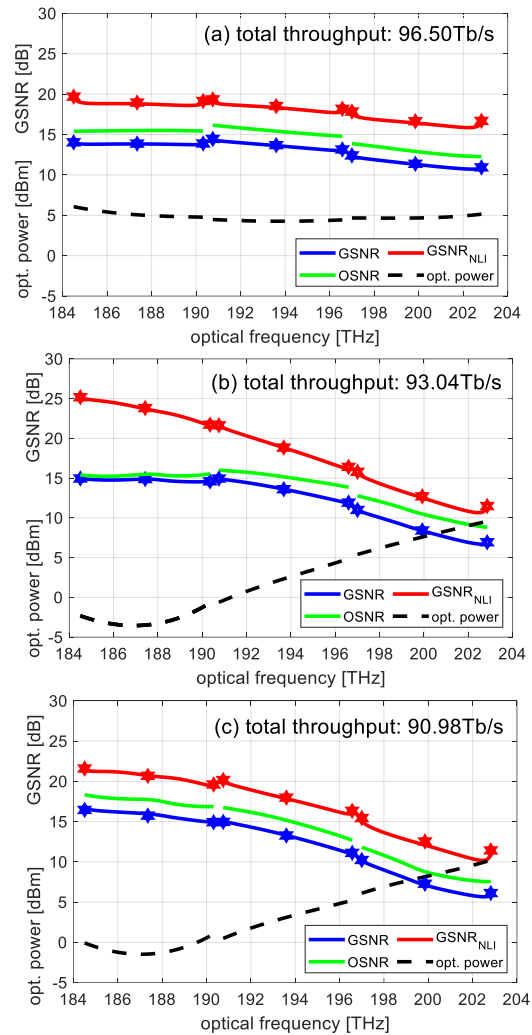

**Fig. 2:** 1000km SMF link, 10x100km, 22.5 dB span loss. (a) **ISRS off**, max throughput optimization. (b) **ISRS on**, max throughput optimization. (c) **ISRS on**, 3-dB rule enforced. Markers: results of numerically integrated EGN-model [10].

The reason for the strong departure from the 3-dB rule, when ISRS is turned on, is that the derivation of the rule hinges on $\eta$ in (3) being independent of launch power. However, when ISRS is turned on, $\eta$ becomes itself a function of launched power, due to Raman transferring power among the channels as they propagate.

Despite the very different look of Fig.2(b) vs. Fig.2(a), the total throughput achieved with ISRS on is only *marginally* lower than that of the system with ISRS off: 93.04 vs. 96.50 Tb/s, respectively (-3.6%). This result qualitatively agrees with the findings of [4] and shows that, by means of careful launch power optimization, the potential performance loss due to the presence of even very strong ISRS can be almost completely mitigated. However, we find that, at the optimum launch power, *low-frequency channels* propagate in linearity, whereas [4] finds lower-frequency channels propagating more non-linearly than (1). Conversely, we find *high-frequency channels* propagating more non-linearly than (1), whereas [4] finds the opposite. We attribute this discrepancy to, possibly, some assumptions in [4] such as a triangular Raman gain profile, strictly Nyquist channel spacing with no inter-band gaps and perhaps the assumption of a uniform 4.5 dB noise-figure across all bands.

The very different look of Fig.2(b) vs. Fig.2(a), would suggest that trying to *force* the 3-dB rule onto that system would likely result in a substantially suboptimal performance. We checked whether that was the case. Fig.2(c) shows an almost perfect compliance with the rule, the red curve being almost exactly 3dB away from the green one throughout. Surprisingly, though, we found a very mild loss of throughput, down to 90.98 Tb/s, only 2.2% lower than Fig.2(b). On the other hand, the peak-to-peak GSNR swing in Fig.2(c) is 11dB, vs. 8dB in Fig.2(b), suggesting that the 3-dB rule has the downside of increasing GSNR non-uniformity.

### 300 km C+L+S+E systems

We then looked at shorter systems (300km), where we considered a larger bandwidth, which now extends partially into the E-band, up to 209.07 THz. Fig.3(a) shows launch power optimization for maximum throughput with ISRS on. Similar to the previous 1000km C+L+S system, the high-frequency channels (now in E-band) need a higher launch power and propagate in a rather non-linear regime. L and C-bands are instead launched at very low power and propagate in linearity. Once more, our results go opposite [4], where a comparable bandwidth case still shows lower non-linearity than (1) for high-frequency channels and higher non-linearity for lower-frequency channels. Similar to [4], launch power optimization for maximum throughput limits the loss from ISRS off to ISRS on, to only 4.6%: 161.40 Tb/s (plot not shown for lack of space) vs. 153.95 Tb/s of Fig.3(a).

We then imposed the 3-dB rule with ISRS on Fig. 3(b). Throughput went down, but only by 2.9%, to 149.37 Tb/s. GSNR peak-to-peak swing went up from 11.1 dB to 15.0 dB, though, confirming such drawback of the 3-dB rule.

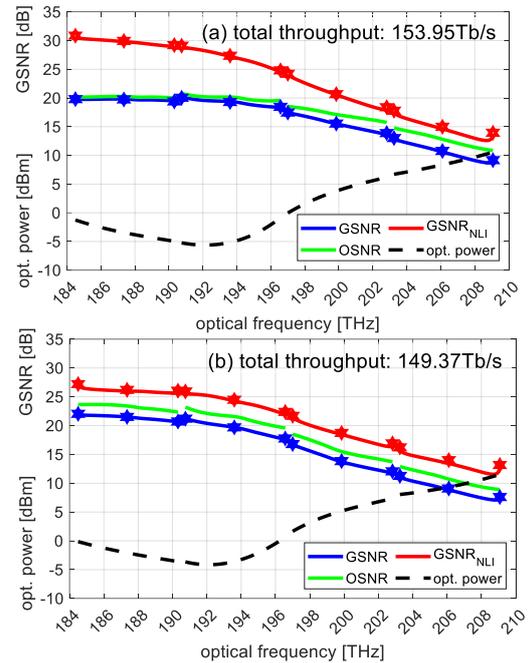

**Fig. 3:** 300km (3x100km) SMF link, **ISRS on.** (a): max throughput optimization. (b): 3-dB rule enforced.

### Discussion and conclusions

We have investigated the penalty due to ISRS on a 300km 24THz C+L+S+E system and a 1000km 18THz C+L+S system, where we tried to make realistic system assumptions. We performed launch power optimization to mitigate the impact of ISRS, using a closed-form multiband physical layer model [5],[6]. We also revisited the "3-dB rule", which prescribes that a gap of 3dB be imposed between $P_{ASE}$ and $P_{NLI}$ (or, equivalently, between NLI-only GSNR and ASE-only OSNR). We checked whether it would hold up in complex multiband systems characterized by frequency-variable parameters and strong ISRS.

Our results indicate that the throughput penalty due to ISRS is surprisingly mild, even in very broadband systems, such those considered in this paper, amounting to less than 5% when launch power optimization is performed. This optimum corresponds to quite different propagation conditions than the 3-dB rule. Nonetheless, in the two study-cases, we found that imposing the 3-dB rule causes less than 3% throughput loss vs. the optimum launch conditions. On the other hand, we also found that the 3-dB rule increases GSNR non-uniformity across channels and therefore its use should be considered with care. Our results broadly agree with those of [4] but the propagation regimes resulting from launch power optimization are different, possibly because of more idealized system assumptions made in [4].

**Acknowledgements**
This work was partially supported by: Cisco Systems through the RISE-ONE Sponsored Research Agreement (SRA); the PhotoNext Center of Politecnico di Torino; the European Union under the Italian National Recovery and Resilience Plan (NRRP) of NextGenerationEU, partnership on "Telecommunications of the Future" (PE00000001 - program "RESTART").